\documentclass[11pt]{article}
\usepackage{cite}
\usepackage{amsmath,amsfonts,amssymb,calrsfs}
\usepackage[small,bf,hang]{caption}
\usepackage{slashed}
\usepackage{mathabx,amsmath}
\usepackage{booktabs}
\usepackage[vcentermath]{youngtab}

\usepackage{stmaryrd}

\def\hybrid{
        \topmargin -20pt
        \oddsidemargin 0pt
        \headheight 0pt \headsep 0pt
        \textwidth 6.25in 
        \textheight 9.5in 
        \marginparwidth .875in
        \parskip 5pt plus 1pt \jot = 1.5ex}

\hybrid
\usepackage{tikz}

 \csname
@addtoreset\endcsname{equation}{section}

\usepackage{color}
\definecolor{red}{rgb}{1,0,0}
\definecolor{lred}{rgb}{0.3,0,0}
\definecolor{green}{rgb}{0,0.6,0}
\definecolor{blue}{rgb}{0,0,1}
\definecolor{violet}{rgb}{0.8,0,0.8}
\definecolor{amber}{rgb}{1.0, 0.75, 0.0}
\definecolor{yellow}{rgb}{1.0, 1.0, 0.0}
\definecolor{applegreen}{rgb}{0.55, 0.71, 0.0}
\definecolor{cadmiumgreen}{rgb}{0.0, 0.42, 0.24}
\definecolor{ballblue}{rgb}{0.13, 0.67, 0.8}
\definecolor{caribbeangreen}{rgb}{0.0, 0.8, 0.6}
\definecolor{bluemunsell}{rgb}{0.0, 0.5, 0.69}
\definecolor{brightpink}{rgb}{1.0, 0.0, 0.5}

\def\moth{\mathsurround=0pt}
\newdimen\zo \zo=0pt

\def\tick{\leaders\hrule height 0.5ex depth 0pt \hskip 0.5pt}
\def\upboxfill{$\moth \setbox\zo\hbox{\tick}%
  \hskip 3pt\hbox to 0pt{$\tick$\hss}\hrulefill \hbox to 7.5pt{$\tick$\hss}$}

\def\dtick{\leaders\hrule height .34pt depth 0.5ex \hskip 0.5pt}
\def\downboxfill{$\moth \setbox\zo\hbox{\dtick}%
  \hskip 2pt\hbox to 0pt{$\dtick$\hss}\hrulefill \hbox to 2pt{$\dtick$\hss}$}

\def\bec{\begin{center}}
\def\ec{\end{center}}

\def\be{\begin{equation}}
\def\ee{\end{equation}}
\def\bea{\begin{eqnarray}}
\def\eea{\end{eqnarray}}
\def\ba{\begin{array}}
\def\ea{\end{array}}

\begin{document}

\begin{titlepage}
\rightline{}
\rightline{June 2022}
\rightline{HU-EP-22/21}
\rightline{LYCEN 2022-01}
\begin{center}
\vskip 1.5cm
 {\LARGE \bf{ 
 Supersymmetric action for 6D $(4,0)$ supergravity } }
\vskip 1.7cm

{\large\bf {Yannick Bertrand$^a$, Stefan Hohenegger$^a$, Olaf Hohm$^b$, Henning Samtleben$^{c, d}$}}
\vskip .8cm

{\it  $^a$ Univ Lyon, Univ Claude Bernard Lyon 1, CNRS/IN2P3, \\
IP2I Lyon, UMR 5822, F-69622, Villeurbanne, France}\\
y.bertrand@ipnl.in2p3.fr, s.hohenegger@ipnl.in2p3.fr
\vskip .3cm

{\it  $^b$ Institute for Physics, Humboldt University Berlin,\\
 Zum Gro\ss en Windkanal 6, D-12489 Berlin, Germany}\\
 ohohm@physik.hu-berlin.de
\vskip .3cm

{\it  $^c$ ENSL, CNRS, Laboratoire de physique, F-69342 Lyon, France} \\
{henning.samtleben@ens-lyon.fr}
\vskip .3cm

{\it  $^d$ Institut Universitaire de France (IUF)}
\vskip .3cm

\vskip .2cm

\end{center}

\bigskip\bigskip
\begin{center} 
\textbf{Abstract}

\end{center} 
\begin{quote}

We give a linearized but otherwise 
complete supersymmetric action for ${\cal N}=(4,0)$ supergravity in six dimensions, 
using a Kaluza-Klein-type $5+1$ split of coordinates and fields.  
We provide in particular a significantly simplified version of the bosonic action derived 
by us recently. This formulation employs fields that are no longer irreducible, subject to a 
local Lorentz invariance, which  in turn simplifies the supersymmetry transformations 
including the exotic gravitino.

\end{quote} 
\vfill
\setcounter{footnote}{0}
\end{titlepage}



\newpage

\section{Introduction}

Maximally supersymmetric theories play a unique role within string and M-theory, with 
their interactions being  completely determined by supersymmetry. 
The unique 11-dimensional supergravity gives rise, upon dimensional reduction, to (ungauged) 
maximal supergravity in lower dimensions, as does the inequivalent type IIB supergravity in 
ten dimensions. Remarkably, however, in six dimensions (6D) there is the possibility of entirely new maximal 
supergravity theories, with multiplets carrying $(3,1)$ and $(4,0)$ supersymmetry \cite{Strathdee:1986jr,Hull:2000zn,Hull:2000rr,Chiodaroli:2011pp,Anastasiou:2013hba,Gunaydin:2020mod
}. 
These multiplets  do not feature a conventional graviton, but instead are based on exotic tensor fields, 
with mixed Young tableaux representations 
${\tiny \yng(2,2)}$ and ${\tiny \yng(2,1)}$, respectively, subject to (exotic) self-duality relations. 
It was conjectured by Hull that these theories are strong coupling limits of ${\cal N}=8$ theories 
in five dimensions \cite{Hull:2000zn}, but it remains unknown whether such exotic theories  actually exist at the interacting level.

Recently, we constructed action principles for the free bosonic parts of the 
$(3,1)$ and $(4,0)$ theories \cite{Bertrand:2020nob}. Since the dynamics of the exotic tensor fields 
is encoded in self-duality relations, it is not possible to write conventional manifestly Lorentz invariant 
actions. However, borrowing techniques from exceptional field theory \cite{Hohm:2013pua,Hohm:2013vpa} 
one may write such actions upon 
performing a $5+1$ split of 
the 6D coordinates (as in Kaluza-Klein theory, but  {without} truncation), 
thereby abandoning manifest Lorentz invariance. 
These  actions take the 
structural form of a 5D theory, but with the fields depending on all six coordinates and with explicit 
contributions from the derivative $\partial_6$ along the sixth dimension. 
The resulting field equations  are in fact  equivalent to  the 
6D self-duality relations. Furthermore, these actions seem to be the ideal starting point in order 
to explore possible interactions, because  in the limit $\partial_6\rightarrow 0$ the interacting  theory 
is immediate, being given by familiar 5D supergravity.\footnote{For this property it is important that 
these actions are, in a precise technical sense \cite{Bertrand:2020nob}, dual to those presented by 
Henneaux-Teitelboim for the case of self-dual $p$-forms \cite{Henneaux:1988gg}. 
See also \cite{Henneaux:2017xsb} for an action principle in the alternative pre-potential formalism.}
A complete action for, say, $(4,0)$ supergravity could then potentially  be constructed as a deformation 
in  $\partial_6$, and there is no longer a reason of principle 
why a consistently interacting theory  should not exist.

In this paper we revisit our formulation for the free $(4,0)$ theory and complete it by introducing 
the fermionic fields, which includes exotic gravitini $\psi_{\mu\nu}^A$ (with USp$(8)$ index $A=1,\ldots, 8$),  
and by constructing  the complete supersymmetric action and  the supersymmetry rules. To this end we give a significant 
simplification of the bosonic action derived in  \cite{Bertrand:2020nob} by trading  the fields that originally live 
in irreducible Young tableaux representations  for reducible fields  subject to 
a local Lorentz invariance. These reducible fields further  absorb certain extra fields, 
which emerged due to the integration of two-derivative self-duality relations in the 5+1 split.

The rest of this paper is organized as follows. In sec.~2 we give the simplified formulation of the bosonic sector 
of the free $(4,0)$ theory. The complete supersymmetric action is then given in sec.~3. 
In sec.~4 we show that on-shell this action indeed implies  the complete 6D supersymmetric $(4,0)$ theory at the 
free level, and we also give, for the first time, the explicit 6D supersymmetry rules.  
We close with a brief outlook, while an appendix collects various gamma matrix and spinor identities.

\section{Bosonic sector}

In this section we review (and improve) the action for the bosonic sector of $(4,0)$ supergravity in 6D. 
To this end we subject the self-duality equations in 6D  to a $5+1$ split of coordinates, 
without truncation, and integrate the result to an action with manifest 5D Lorentz invariance. 
We also provide a frame formulation with local Lorentz  invariance, which in turn 
is convenient for the formulation of the supersymmetry rules provided in later sections.

\subsection{6D self-duality relations  and 5+1 split}
\label{subsec:split40}

The ${\cal N}=(4,0)$ multiplet in 6D contains 42 scalar fields, 27 self-dual two-forms, and a self-dual four-index tensor. We begin 
with the self-dual two-forms $B_{\hat\mu\hat\nu}{}^M$, labelled by an index $M=1, \dots, 27$, with abelian field strength 
$H_{\hat\mu\hat\nu\hat\rho}{}^M =3\,\partial_{[\hat\mu}  B_{\hat\nu\hat\rho]}{}^M$, 
where all hatted indices are 6D spacetime indices. The field strength obeys  the
self-duality equations 
\bea
H_{\hat\mu\hat\nu\hat\rho}{}^M &=& 
\frac16\,\varepsilon_{\hat\mu\hat\nu\hat\rho\hat\sigma\hat\kappa\hat\lambda}\,H^{\hat\sigma\hat\kappa\hat\lambda}{}^M
\;. 
\label{6DsdB}
\eea
These equations cannot be obtained from a manifestly Lorentz invariant action in 6D, 
but an action can be provided upon performing a 5+1 split of coordinates 
(hence abandoning manifest Lorentz invariance). Specifically, we decompose the coordinates and indices as 
\bea
\{ x^{\hat{\mu}}\} \longrightarrow \{x^\mu, x^6=y \} 
\;,\qquad
\mu=0, 1, \dots, 4\;,
\label{split51}
\eea 
and similarly the  6D tensor fields as 
 \bea
  \{B_{\hat\mu\hat\nu}{}^M{}\}=\{B_{\mu\nu}{}^M{}, B_{\mu 6}{}^M{} \equiv A_\mu{}^M\}\,. 
  \label{AB6}
 \eea
As shown in \cite{Bertrand:2020nob}, the components of 
the self-duality equations (\ref{6DsdB})  
are then equivalently expressed by the duality equation
\bea
{\cal F}_{\mu\nu}{}^M + \frac16 \varepsilon_{\mu\nu\rho\sigma\tau}\,H^{\rho\sigma\tau\,M} &=& 0
\;,
\label{eomFH}
\eea
which in turn can  be obtained by variation of the Lagrangian
\bea
{\cal L}_B &=& 
-\frac12\,{\cal F}_{\mu\nu}{}^M {\cal F}^{\mu\nu}{}^M
-\frac1{12}\,\varepsilon^{\mu\nu\rho\sigma\tau}\,\partial_6 B_{\mu\nu}{}^M\,H_{\rho\sigma\tau}{}^M
\;,
\label{ExFTsd}
\eea
where $H_{\mu\nu\rho}{}^M=3\,\partial_{[\mu} B_{\nu\rho]}{}^M$, and  we defined 
\bea
{\cal F}_{\mu\nu}{}^M \ \equiv \ {F}_{\mu\nu}{}^M+\partial_6 B_{\mu\nu}{}^M
\ \equiv \ 2\partial_{[\mu} A_{\nu]}{}^M +\partial_6 B_{\mu\nu}{}^M
\;.
\label{FB}
\eea
Note that this Lagrangian is invariant under gauge transformations of 
the structural form of a tensor hierarchy: 
 \be
  \delta A_{\mu}{}^M  \ =  \ \partial_{\mu}\Lambda{}^M - \partial_6 \Xi_{\mu}{}^M\;, \qquad
  \delta B_{\mu\nu}{}^M \ = \ 2\,\partial_{[\mu}\Xi_{\nu]}{}^M + \chi_{\mu\nu}{}^M\;, 
  \label{deltaAB}
 \ee
where $\partial_6\chi_{\mu\nu}{}^M=0$.  The $\chi$ gauge invariance follows from the fact that this shift in $B_{\mu\nu}{}^M$
changes the Lagrangian only by a total $\partial_6$ derivative. 

We next mimic this strategy for the exotic four-index tensor field contained in 
the ${\cal N}=(4,0)$ multiplet. This field is denoted by $T_{\hat\mu\hat\nu,\hat\rho\hat\sigma}$
and carries the symmetries of the (2,2) window Young tableaux (equivalently, the symmetries of the Riemann tensor): 
\bea
{\tiny \yng(2,2)} &:&
T_{\hat\mu\hat\nu,\hat\rho\hat\sigma}=T_{\hat\rho\hat\sigma,\hat\mu\hat\nu}=-T_{\hat\nu\hat\mu,\hat\rho\hat\sigma}\;,\quad
T_{[\hat\mu\hat\nu,\hat\rho]\hat\sigma}= 0
\;.
\label{T40}
\eea 
This field admits a curvature tensor of second order in derivatives:
\bea
G_{\hat{\mu}\hat{\nu}\hat{\lambda},\hat{\rho}\hat{\sigma}\hat{\tau}} &=& 
3\, \partial_{\hat{\rho}} \partial_{[\hat{\mu}}T_{\hat{\nu}\hat{\lambda}],\hat{\sigma}\hat{\tau}}
+3\, \partial_{\hat{\sigma}} \partial_{[\hat{\mu}}T_{\hat{\nu}\hat{\lambda}],\hat{\tau}\hat{\rho}}
+3\, \partial_{\hat{\tau}} \partial_{[\hat{\mu}}T_{\hat{\nu}\hat{\lambda}],\hat{\rho}\hat{\sigma}}
 \;, 
 \label{defGT}
\eea
which is invariant under the gauge transformations 
\bea
\delta T_{\hat{\mu}\hat{\nu},\hat{\rho}\hat{\sigma}} &=& 
\partial_{[\hat{\mu}} \lambda_{\hat{\nu}],\hat{\rho}\hat{\sigma}} + 
\partial_{[\hat{\rho}} \lambda_{\hat{\sigma}],\hat{\mu}\hat{\nu}} 
\;,
\label{gauge40}
\eea
with a gauge parameter obeying $\lambda_{\hat{\mu},\hat{\rho}\hat{\sigma}}=\lambda_{\hat{\mu},[\hat{\rho}\hat{\sigma}]}$, 
$\lambda_{[\hat{\mu},\hat{\rho}\hat{\sigma}]}=0$ (and hence living in the $(2,1)$ Young tableaux ${\tiny \yng(2,1)}$\,). 
The field equations are given by the 6D self-duality relations 
\bea
G_{\hat{\mu}\hat{\nu}\hat{\lambda},\hat{\rho}\hat{\sigma}\hat{\tau}} &=& 
\frac16\,\varepsilon_{\hat{\mu}\hat{\nu}\hat{\lambda}\hat{\alpha}\hat{\beta}\hat{\gamma}}\,
G^{\hat{\alpha}\hat{\beta}\hat{\gamma}}{}_{\hat{\rho}\hat{\sigma}\hat{\tau}}
\;,
\label{6DsdT40}
\eea
which  are manifestly gauge invariant.

Subjecting this exotic tensor field  to the same 5+1 split of coordinates as in (\ref{split51}) 
yields the following components: 
\bea
\{ T_{\hat\mu\hat\nu,\hat\rho\hat\sigma} \} &=&
\left\{ T_{\mu\nu,\rho\sigma}\,;\;T_{\mu\nu,\rho6} = C_{\mu\nu,\rho}\,;\; T_{\mu6,\nu6} = h_{\mu\nu} \right\}
\;.
\label{TCh40}
\eea
We note that, after proper dimensional reduction to five dimensions, these fields describe the graviton, dual graviton and double dual graviton, respectively.
In this parametrization the six-dimensional field equations (\ref{6DsdT40}) split into two equations: 
\bea
R_{\mu\nu,\rho\sigma}
&=&
\frac12\,\partial_6  \partial_{\mu}{C}_{\rho\sigma,\nu}
-\frac12\,\partial_6  \partial_{\nu}{C}_{\rho\sigma,\mu}
+\frac12\, \partial_6 \partial_{\rho} {C}_{\mu\nu,\sigma}
-\frac12\,\partial_6  \partial_{\sigma} {C}_{\mu\nu,\rho}
\nonumber\\
&&{} 
+\frac12\,\varepsilon_{\mu\nu\kappa\lambda\tau}\,
\partial_{[\rho} \partial^{\kappa}{C}^{\lambda\tau}{}_{\sigma]}
+\frac14\,\varepsilon_{\mu\nu\kappa\lambda\tau}\,
\partial_6  \partial^{\kappa}T^{\lambda\tau}{}_{\rho\sigma}
+\frac12\, \partial_6  \partial_6 T_{\mu\nu,\rho\sigma}
\;,\qquad
\label{6Dsd1}\\[2ex]
\varepsilon_{\mu\nu\alpha\beta\gamma}\,
\partial^\alpha \partial_{[\rho} T_{\sigma\tau]}{}^{\beta\gamma}
&=&
- 2\,\partial_{\mu} \partial_{[\rho}{C}_{\sigma\tau],\nu}
+ 2\,\partial_{\nu} \partial_{[\rho}{C}_{\sigma\tau],\mu}
- 2\,\partial_6  \partial_{[\rho}T_{\sigma\tau],\mu\nu}
\;,
\label{6Dsd2}
\eea
with the linearized Riemann tensor $R_{\mu\nu,\rho\sigma}$ 
for the field $h_{\mu\nu}$\,.

These equations can be integrated to a form for which an action can be written. 
The second equation (\ref{6Dsd2}) has the form of a curl in $[\rho\sigma\tau]$ and can hence be integrated to 
\bea
\frac12\,\varepsilon_{\mu\nu\alpha\beta\gamma}\,
\partial^\alpha  T_{\sigma\tau}{}^{\beta\gamma}
+ \partial_{\mu} {C}_{\sigma\tau,\nu}
- \partial_{\nu} {C}_{\sigma\tau,\mu}
+ \partial_6  T_{\sigma\tau,\mu\nu}
&=&
2\, \partial_{[\sigma} v_{\tau],\mu\nu}
\;,
\label{defv}
\eea
up to a tensor $v_{\tau,\mu\nu}=-v_{\tau,\nu\mu}$ that is  determined by this equation up
to the gauge freedom 
\bea
\delta v_{\tau,\mu\nu}&=&\partial_\tau \zeta_{\mu\nu}\;,
\label{delv}
\eea
with $\zeta_{\mu\nu}=-\zeta_{\nu\mu}$.
Combining (\ref{defv}) with the first field equation (\ref{6Dsd1}), we find
\bea
R_{\mu\nu,\rho\sigma}
&=&
\frac12\, \partial_6 \partial_{\rho} {C}_{\mu\nu,\sigma}
-\frac12\, \partial_6  \partial_{\sigma} {C}_{\mu\nu,\rho}
+\frac12\, \varepsilon_{\mu\nu\kappa\lambda\tau}\,
\partial_{[\rho} \partial^{\kappa} {C}^{\lambda\tau}{}_{\sigma]}
+\partial_6 \partial_{[\rho} v_{\sigma],\mu\nu}
\;,
\label{6Dsd1A}
\eea
which in turn is a curl in $[\rho\sigma]$ and can hence be integrated to 
\bea
\partial_{[\mu} h_{\nu]\rho}
+\frac1{4}\,\varepsilon_{\mu\nu\lambda\sigma\tau}\,
 \partial^{\lambda}{C}^{\sigma\tau}{}_{\rho}
+ \frac1{2}\,\partial_6  {C}_{\mu\nu,\rho}
+\frac12\,\partial_6  v_{\rho,\mu\nu}
&=&
\partial_\rho u_{\mu\nu}
\;,
\label{defu40}
\eea
up to an antisymmetric field $u_{\mu\nu}=-u_{\nu\mu}$\,.
In \cite{Bertrand:2020nob}, we have given a Lagrangian (with manifest 5D Lorentz invariance)
whose field equations are precisely equivalent to equations (\ref{defv}) and (\ref{defu40}).
We do not repeat its somewhat lengthy form here as it shall be superseded by the construction in 
sections~\ref{subsec:field_redefinitions}, \ref{subsec:frame} below.

The gauge symmetries of equations (\ref{defv}) and (\ref{defu40}) originate 
from the original 6D gauge transformations (\ref{gauge40}).
With the decomposition
\bea
\left\{
\lambda_{\hat\rho,\hat\mu\hat\nu}
\right\} &=&
\left\{
\lambda_{\rho,\mu\nu}\,;\;
\lambda_{\mu,\nu 6}=2\,\alpha_{\mu\nu}-\frac23\,\beta_{\mu\nu}\,;\;
\lambda_{6,\mu6} = 2\,\xi_\mu
\right\} 
\;,
\eea
with symmetric $\alpha_{\mu\nu}$, and antisymmetric $\beta_{\mu\nu}$,
their action on the various components of (\ref{TCh40}) is found to be
\bea
\delta h_{\mu\nu} &=& 
2\,\partial_{(\mu} \xi_{\nu)}
-2\,\partial_6  {\alpha}_{\mu\nu}
\;,\nonumber\\
\delta {C}_{\mu\nu,\rho} &=&
2\,\partial_{[\mu} \alpha_{\nu]\rho}  
+\partial_\rho{\beta}_{\mu\nu}
-\partial_{[\rho}{\beta}_{\mu\nu]}
-\frac12\,\partial_6   \lambda_{\rho,\mu\nu} 
\;,
\nonumber\\
\delta T_{\mu\nu,\rho\sigma} &=& 
\partial_{[\mu} \lambda_{\nu],\rho\sigma} + 
\partial_{[\rho} \lambda_{\sigma],\mu\nu} 
\;.
\label{gauge40_1}
\eea
The gauge variations of the two new fields $v_{\rho,\mu\nu}$ and $u_{\mu\nu}$ are obtained
by integrating the variation of (\ref{defv}) and (\ref{defu40}), respectively, giving rise to
\bea
\delta u_{\mu\nu}
&=&
 \partial_{[\mu} \xi_{\nu]}
+ \frac1{6}\, \varepsilon_{\mu\nu\lambda\sigma\tau}\partial^{\lambda} \beta^{\sigma\tau} 
+ \frac1{3}\,\partial_6  \beta_{\mu\nu} 
+ \frac1{2}\,\partial_6  \zeta_{\mu\nu} 
\;,
\nonumber\\
 \delta v_{\rho,\mu\nu}
&=&
\frac14\,\varepsilon_{\mu\nu\kappa\lambda\sigma}\,
\partial^\kappa   \lambda_{\rho}{}^{\lambda\sigma}  
+2\,\partial_{[\mu} \alpha_{\nu]\rho} 
+\frac23\,\partial_{[\mu} \beta_{\nu]\rho}
+\partial_\rho\,\zeta_{\mu\nu}
+\frac{1}{2}\, \partial_6   \lambda_{\rho,\mu\nu}
\;,
\label{symmetries40}
\eea
with $\zeta_{\mu\nu}=-\zeta_{\nu\mu}$ introduced in (\ref{delv}).

\subsection{Field redefinitions }
\label{subsec:field_redefinitions}

Next we perform field redefinitions that simplify the structure of the theory. The above  fields 
$(h_{\mu\nu}, C_{\mu\nu,\rho}, T_{\mu\nu,\rho\sigma})$ 
live in {irreducible} Young tableau representations, but  it turns out to be beneficial to 
reorganize them into {reducible} representations, which in particular absorbs the newly 
introduced fields $u_{\mu\nu}, v_{\rho,\mu\nu}$ but also simplifies  the 
gauge transformations.\footnote{In related context, the use of reducible representations has also allowed a uniform formulation of the known linearized supergravities~\cite{Sorokin:2018djm}.}
We begin with the four-index tensor $T_{\mu\nu,\rho\sigma}$ which we replace by 
\bea
{{\cal T}}_{\mu\nu,\rho\sigma} &:=& 
T_{\mu\nu,\rho\sigma} +2\,(\eta_{\mu[\rho} h_{\sigma]\nu}-\eta_{\nu[\rho} h_{\sigma]\mu})
 -4\,  \eta_{\mu[\rho}\,u_{\sigma]\nu} 
 +4\,  \eta_{\nu[\rho}\,u_{\sigma]\mu} 
\nonumber\\
&&{}
+\varepsilon_{\rho\sigma\kappa\lambda[\mu}\,  {{C}}^{\kappa\lambda}{}_{\nu]}  
-\varepsilon_{\rho\sigma\kappa\lambda[\mu}\, {v}_{\nu]}{}^{\kappa\lambda}
\;.
\label{newT}
\eea
The above first-order field equations (\ref{defv}), (\ref{defu40}) then imply that
\bea
3\,\partial_{[\mu} {{\cal T}}_{\nu\rho],\sigma\tau}
&=&
\partial_6 {\cal J}_{\mu\nu\rho,\sigma\tau}
\;,
\label{eomTT}
\eea
with the current
\bea
{\cal J}_{\mu\nu\rho,\sigma\tau} &=&
\frac32\,\varepsilon_{\sigma\tau \kappa\lambda [\mu} {\cal T}_{\nu\rho]}{}^{\kappa\lambda} 
+6\,  \eta_{\sigma[\mu}    \widehat{C}_{\nu\rho],\tau} 
-6\,  \eta_{\tau[\mu}    \widehat{C}_{\nu\rho],\sigma} 
-6\,\varepsilon_{\sigma\tau \lambda [\mu\nu}  h_{\rho]}{}^{\lambda} 
\;,
\label{currentJ}
\eea
where we have also defined the reducible field
\bea
\widehat{C}_{\mu\nu,\rho} &=& C_{\mu\nu,\rho} + \varepsilon_{\mu\nu\rho\sigma\tau} u^{\sigma\tau}
\;.
\label{newC}
\eea

The gauge transformations (\ref{symmetries40}) for these reducible fields can similarly   be brought  to a form 
with reducible gauge parameters: 
\bea
\delta h_{\mu\nu} &=&
 2\,\partial_{(\mu} \xi_{\nu)}
  -2\,\partial_6 \Gamma_{(\mu\nu)}\,,
\nonumber\\
\delta \widehat{C}_{\mu\nu,\rho} &=&
2\,\partial_{[\mu} \Gamma_{\nu]\rho}  
+ \varepsilon_{\mu\nu\rho\sigma\tau} \partial^\sigma \xi^\tau
+ {\partial_6}  \Lambda_{[\mu,\nu]\rho} 
- \varepsilon_{\mu\nu\rho\sigma\tau} {\partial_6}  \Gamma^{\sigma\tau} 
-2\,{  \eta_{\rho[\mu} \partial_6} \xi_{\nu]} 
\;,
\nonumber\\
  \delta {{\cal T}}_{\mu\nu,\rho\sigma} &=& 
  2\,\partial_{[\mu} \Lambda_{\nu],\rho\sigma} - 
 \varepsilon_{\rho\sigma\kappa\lambda[\mu}\,
 \partial_6 \Lambda_{\nu]}{}^{\kappa\lambda} 
   -4\,\eta_{\rho[\mu}  \partial_6  \Gamma_{\nu]\sigma}
   +4\,\eta_{\sigma[\mu}  \partial_6  \Gamma_{\nu]\rho}
-4\,  \varepsilon_{\mu\nu\rho\sigma\tau}\,  \partial_6  \xi^{\tau}  
\,,  \qquad 
\label{gauge_new}
\eea
where the reducible gauge parameters $\Gamma_{\mu\nu}$, $\Lambda_{\nu,\rho\sigma}$ 
are related to the original ones (\ref{gauge40_1}) via 
\bea
\Gamma_{\mu\nu} &=& \alpha_{\mu\nu}-\beta_{\mu\nu}
\;,
\nonumber\\
  \Lambda_{\nu,\rho\sigma} &=& \lambda_{\nu,\rho\sigma} 
  +\frac12\,\varepsilon_{\nu\rho\sigma\kappa\lambda}\,\zeta^{\kappa\lambda} 
  -\frac23\,\varepsilon_{\nu\rho\sigma\kappa\lambda}\,\beta^{\kappa\lambda} 
  -4\,\eta_{\nu[\rho}\,\xi_{\sigma]} 
  \;.
\eea
The field equations (\ref{eomTT}) show that
after reduction to 5D (i.e.\ in the limit $\partial_6\rightarrow0$), the field ${{\cal T}}_{\mu\nu,\rho\sigma}$ 
satisfies $3\,\partial_{[\mu} {{\cal T}}_{\nu\rho],\sigma\tau}=0$, such that it can 
be gauged to zero by the gauge transformations (\ref{gauge_new}), which in this limit reduce to
\bea
  \delta {{\cal T}}_{\mu\nu,\rho\sigma} &=& 
  2\,\partial_{[\mu} \Lambda_{\nu],\rho\sigma} 
  \;.
\label{gaugeT0}
\eea
This is another manifestation of the fact that in 5D, after gauge fixing,
the double dual graviton is related to the graviton by a purely algebraic relation~\cite{Hull:2000rr} (see also \cite{Henneaux:2019zod}).

In terms of these new field variables we can finally give the Lagrangian that reproduces 
the above integrated field equations and that is equivalent to the Lagrangian  given in 
 \cite{Bertrand:2020nob}. Specifically, for the fields 
$(h_{\mu\nu}, \widehat{C}_{\mu\nu,\rho}, {\cal T}_{\mu\nu,\rho\sigma})$, 
which here and in the following we refer to as  the spin-2 sector, the Lagrangian reads 
\bea
{\cal L} &=&
-\frac{1}{4}\,\widehat{\Omega}^{\mu\nu\rho} \widehat{\Omega}_{\mu\nu\rho}
+\frac12\,\widehat{\Omega}^{\mu\nu\rho}\widehat{\Omega}_{\nu\rho\mu}
+\widehat{\Omega}^{\mu}\widehat{\Omega}_{\mu}
\nonumber\\
&&{}
- \frac1{8}\,\varepsilon^{\mu\nu\rho\sigma\tau}\,
\partial_{\mu} \widehat{{C}}{}_{\nu\rho,\alpha} \,\partial_6  \widehat{{C}}{}_{\sigma\tau}{}^\alpha
-\frac1{64}\,
\varepsilon^{\mu\nu\rho\sigma\tau} 
\partial_\mu {\cal T}_{\nu\rho,\alpha\beta}
\partial_6 {\cal T}_{\sigma\tau}{}^{\alpha\beta} 
\nonumber\\
&&{}
  +\frac{3}{4}\,
\partial_6 h_{\mu}{}^{\sigma}
\, \partial_6 h_{\sigma}{}^{\mu} 
-\frac{3}{4}\,
\partial_6 h_{\mu}{}^{\mu}
\, \partial_6 h_\nu{}^\nu
+
\frac1{96}\,
\varepsilon^{\mu\nu\rho\sigma\tau} 
\partial_6 {\cal T}_{\mu\nu,\alpha\beta}
 \partial_6 {\cal J}'_{\rho\sigma\tau}{}^{\alpha\beta} \;, 
 \label{L_spin2}
\eea
where we defined 
\bea
\widehat{\Omega}_{\mu\nu\rho} &=& \partial_{[\mu} h_{\nu]\rho}
+\partial_6 \widehat{C}_{\mu\nu,\rho}
\;,\nonumber\\
{\cal J}'_{\mu\nu\rho,\sigma\tau} &=&
{\cal J}_{\mu\nu\rho,\sigma\tau}-
\frac34\,\varepsilon_{\sigma\tau \kappa\lambda [\mu} {\cal T}_{\nu\rho]}{}^{\kappa\lambda} 
\;.
\label{Omegahat}
\eea
We have verified that this  reproduces the Lagrangian given in \cite{Bertrand:2020nob}.
Note that $\widehat{C}_{\mu\nu,\rho}$ and ${\cal T}_{\mu\nu,\rho\sigma}$ appear in the Lagrangian 
only under $\partial_6$ derivatives (up to total derivatives), which implies a gauge invariance under shifts that depend only on 5D 
coordinates. This gauge invariance allows one to gauge to zero the `integration constants' obtained by 
integrating an equation of motion that is a total  $\partial_6$ derivative. 
Using this, variation w.r.t.\ ${\cal T}_{\mu\nu,\rho\sigma}$ directly yields the equations (\ref{eomTT}).
Let us also note the field equation obtained by 
variation w.r.t.~$\widehat{C}_{\mu\nu,\rho}$, 
\bea
 \frac1{4}\,\varepsilon_{\mu\nu\lambda\sigma\tau}\,\partial^{\lambda} \widehat{{C}}{}^{\sigma\tau}{}_{\rho}
+\frac{1}{2}\,\widehat{\Omega}_{\mu\nu\rho} 
-\widehat{\Omega}_{\rho[\mu\nu]}  
-2\,\widehat{\Omega}_{[\mu} \,\eta_{\nu]\rho}
&=&
\frac{1}{8}\,
\varepsilon_{\mu\nu\sigma\tau\lambda} 
\partial_6 {\cal T}^{\sigma\tau,\lambda}{}_{\rho}{}
   \;, 
   \label{eomCC}
\eea
which reproduces some linear combination of the above duality equations (\ref{defv}), (\ref{defu40}).

\subsection{Frame formulation}
\label{subsec:frame}

In this final subsection we give a further generalization of the above theory, which turns out to be beneficial, 
in particular for the presentation of the supersymmetry variations. We give a frame formulation with a local 
Lorentz invariance, for which the `metric fluctuations' $h_{\mu\nu}$ receive an antisymmetric part, 
thereby rendering also this field reducible. Concretely, 
we postulate that the two fields $(h_{\mu\nu}, \widehat{C}_{\mu\nu,\rho})$ 
are replaced by reducible fields $(e_\mu{}^\alpha$, ${\cal C}_{\mu\nu}{}^\alpha)$ with gauge transformations
\bea
\delta e_{\mu}{}^\alpha &=&
 \partial_{\mu} \xi^\alpha 
  -\partial_6 \Gamma_{\mu}{}^\alpha -\Lambda_{\mu}{}^\alpha 
  \;,
\nonumber\\
\delta {\cal C}_{\mu\nu,\alpha}  &=&
2\,\partial_{[\mu} \Gamma_{\nu]\alpha}    
+ \varepsilon_{\mu\nu\alpha\beta\gamma} \,\Lambda^{\beta\gamma}
+ {\partial_6}  \Lambda_{[\mu\nu],\alpha}  
-2\,{  \delta_{\alpha[\mu} \partial_6} \xi_{\nu]} 
\;,\quad
\label{gauge_wL}
\eea
where we introduced local 
Lorentz transformations with gauge parameter $\Lambda_{\mu\nu}=-\Lambda_{\nu\mu}$.
We should emphasize that the indices $\alpha, \beta, \ldots$ are ordinary spacetime indices, 
on the same footing as $\mu, \nu, \ldots$, but occasionally we shall chose such letters from the beginning of the greek alphabet 
in order to highlight their apparent role as frame indices of a putative non-linear theory of frame fields. 
The gauge transformations of ${\cal T}_{\mu\nu,\rho\sigma}$ remain unchanged w.r.t.\ (\ref{gauge_new}): 
\bea
  \delta {{\cal T}}_{\mu\nu,\rho\sigma} &=& 
  2\,\partial_{[\mu} \Lambda_{\nu],\rho\sigma} - 
 \varepsilon_{\rho\sigma\kappa\lambda[\mu}\,
 \partial_6 \Lambda_{\nu]}{}^{\kappa\lambda} 
   -4\,\eta_{\rho[\mu}  \partial_6  \Gamma_{\nu]\sigma}
   +4\,\eta_{\sigma[\mu}  \partial_6  \Gamma_{\nu]\rho}
-4\,  \varepsilon_{\mu\nu\rho\sigma\tau}\,  \partial_6  \xi^{\tau}  
\,.  \qquad 
\label{deltaT}
\eea

Let us first check that if we had a formulation with gauge symmetry (\ref{gauge_wL}) we could always 
fix a gauge so as to recover the gauge transformations in the form (\ref{gauge_new}). 
There are two natural gauge conditions: First, we can pick the `symmetric' gauge that the antisymmetric 
part of $e$ vanishes, $e_{[\mu\nu]}=0$, in which case the spin-2 fluctuation of the previous subsection 
is recovered as $h_{\mu\nu}=2 e_{\mu\nu}$. In order not to undo the gauge fixing we have 
to add compensating gauge transformations: 
 \be
  \delta e_{[\mu\nu]}=0 \quad \Rightarrow \quad \Lambda_{\mu\nu}=\partial_{[\mu} \xi_{\nu]}-\partial_6\Gamma_{[\mu\nu]}
  \;. 
  \label{e00}
 \ee
Upon identifying  ${\cal C}_{\mu\nu}{}^\alpha=\widehat{C}_{\mu\nu}{}^\alpha$ 
this reproduces precisely the gauge transformation $\delta \widehat{C}_{\mu\nu}{}^\alpha$ in the second line of (\ref{gauge_new}). Second, we may pick the gauge condition that the totally antisymmetric part of 
${\cal C}_{\mu\nu}{}^\alpha$ vanishes 
(thanks to the local Lorentz parameter acting according to the second term in $\delta \widehat{C}_{\mu\nu}{}^\alpha$ in 
(\ref{gauge_new})). In this case, the compensating gauge transformations following from 
  \be
   0= \delta {\cal C}_{[\mu\nu,\rho]}  \quad \Rightarrow \quad
      \Lambda^{\alpha\beta} = \frac{1}{6} \,\varepsilon^{\alpha\beta\mu\nu\rho}\partial_{\mu}\Gamma_{\nu\rho}
    -  \frac{1}{12} \,\varepsilon^{\alpha\beta\mu\nu\rho}\partial_6 \Lambda_{\mu,\nu\rho}\;, 
  \ee 
imply that we can identify 
  \be
    {\cal C}_{\mu\nu,\rho} = {C}_{\mu\nu,\rho}\;,\qquad    
     2\, e_{\mu\nu} = h_{\mu\nu} +2\, u_{\mu\nu}\;, 
  \ee 
as one may verify with (\ref{gauge_new}).

Let us finally write down the Lagrangian that permits to realize the enhanced gauge symmetry (\ref{gauge_wL}). 
We may start from (\ref{L_spin2}) and promote the fields accordingly, but we have to allow for the 
possibility that terms need to be added that vanish in the above gauges. This is indeed what happens, 
and one finds the fully gauge invariant Lagrangian to be given by
\bea
{\cal L} &=&
-\frac{1}{4}\,\widehat{\Omega}^{\mu\nu\rho} \widehat{\Omega}_{\mu\nu\rho}
+\frac12\,\widehat{\Omega}^{\mu\nu\rho}\widehat{\Omega}_{\nu\rho\mu}
+\widehat{\Omega}^{\mu}\widehat{\Omega}_{\mu}
 + \frac{1}{2} \,\varepsilon^{\mu\nu\rho\sigma\tau} \,\partial_{\mu} e_{\nu\rho} \,\partial_6 e_{\sigma\tau} 
\nonumber\\
&&{}
- \frac1{8}\,\varepsilon^{\mu\nu\rho\sigma\tau}\,
\partial_{\mu} {\cal C}{}_{\nu\rho,\alpha} \,\partial_6  {\cal C}{}_{\sigma\tau}{}^\alpha
-\frac1{64}\,
\varepsilon^{\mu\nu\rho\sigma\tau} 
\partial_\mu {\cal T}_{\nu\rho,\alpha\beta}
\partial_6 {\cal T}_{\sigma\tau}{}^{\alpha\beta} 
\nonumber\\
&&{}
 + 3\,
\partial_6 e^{\mu\nu}
\, \partial_6 e_{\nu\mu} 
-3\,
\partial_6 e_{\mu}{}^{\mu}
\, \partial_6 e_\nu{}^\nu
+
\frac1{96}\,
\varepsilon^{\mu\nu\rho\sigma\tau} 
\partial_6 {\cal T}_{\mu\nu,\alpha\beta}
 \partial_6 {\cal J}'_{\rho\sigma\tau}{}^{\alpha\beta} \nonumber\\
 &&{}
  +\frac{1}{2}\, \varepsilon^{\mu\nu\rho\alpha\beta} \partial_6 e_{\alpha\beta} \partial_6 {\cal C}_{\mu\nu,\rho}\,,
  \label{frameinvLag}
\eea
with
\bea
\widehat{\Omega}_{\mu\nu}{}^{\rho} &=& 2\,\partial_{[\mu} e_{\nu]}{}^{\rho} + \partial_6{\cal C}_{\mu\nu}{}^{\rho}
\;,\nonumber\\
{\cal J}'_{\mu\nu\rho}{}^{\sigma\tau} &=&
\frac34\,\varepsilon^{\sigma\tau}{}_{\beta\gamma[\mu}\, {\cal T}_{\nu\rho]}{}^{\beta\gamma} 
+12\,  \delta_{[\mu}{}^{[\sigma}    {\cal C}_{\nu\rho]}{}^{\tau]} 
-12\,\varepsilon^{\sigma\tau}{}_{\gamma[\mu\nu}\,  e_{\rho]}{}^\gamma 
\;.
\label{Omegahat_e}
\eea
Upon choosing the gauge condition 
$e_{[\mu\nu]}=0$ this reduces to the previous Lagrangian (\ref{L_spin2}). 
Thus, the Lagrangian (\ref{frameinvLag})  equivalently defines the spin-2 sector of ${\cal N}=(4,0)$ supergravity. 
Again, only 5D Lorentz invariance is manifest, yet all fields still depend on all six coordinates.

\section{${\cal N}=(4,0)$ supersymmetric Lagrangian in 5+1 split}

In this section, we will start from the bosonic Lagrangian of ${\cal N}=(4,0)$ supergravity
built as a linear combination of (\ref{frameinvLag}) with (\ref{ExFTsd}) and the scalar Lagrangian
and construct its fermionic extension as a maximally supersymmetric theory. 
Similar to the bosonic fields,
the fermions in this formulation take the form of 5D fermions, however depending on all six space-time coordinates.
In the supersymmetric model, all fields transform in representations of the 5D R-symmetry group ${\rm USp}(8)$\,.
While the spin-2 sector is invariant under the R-symmetry, we relabel the 27 vector and tensor fields from (\ref{AB6}) as
\bea
\left\{A_\mu{}^{AB}, B_{\mu\nu}{}^{AB}\right\}
\;,\qquad
A, B = 1, \dots, 8\;,
\eea
by antisymmetric, symplectic traceless, index pairs $[AB]$ of fundamental ${\rm USp}(8)$ indices, i.e.
\bea
A_\mu{}^{AB}=-A_\mu{}^{BA}\;,\quad
A_\mu{}^{AB}\,\Omega_{AB} =0
\;,
\eea
with the antisymmetric ${\rm USp}(8)$ invariant symplectic tensor $\Omega_{AB}$. Similarly, the 42 scalar fields
of the free theory are most conveniently labelled by an antisymmetric traceless product of ${\rm USp}(8)$ indices as
\bea
\phi^{ABCD} = \phi^{\llbracket ABCD \rrbracket}\;,\quad
\mbox{i.e.}\;\;
\phi^{ABCD} = \phi^{[ABCD]}\;\;\mbox{and}\;\;
\phi^{ABCD}\,\Omega_{CD}=0
\;.
\eea

The fermionic sector of the model contains the 8 gravitinos $\psi_\mu^A$ in the fundamental 
representations of the 5D R-symmetry group ${\rm USp}(8)$, which we expect in maximal 5D supergravity after
reduction to five dimensions. Similarly, the model carries the 48  spin 1/2 fields of maximal 5D supergravity, labelled as
\bea
\chi^{ABC} = \chi^{\llbracket ABC\rrbracket}\;,\qquad
\chi^{ABC}\,\Omega_{BC} = 0
\;. 
\eea
We refer to \cite{deWit:2004nw} for our conventions of 5D USp(8) spinors.
In particular, we have
\bea
\gamma^{\mu\nu\rho\sigma\tau} = -i\varepsilon^{\mu\nu\rho\sigma\tau}
\;,\quad
\bar\epsilon\chi = \bar\chi\epsilon\;,\quad
\bar\epsilon\gamma_\mu \chi = \bar\chi\gamma_\mu \epsilon\;,\quad
\bar\epsilon\gamma_{\mu\nu}\chi = -\bar\chi\gamma_{\mu\nu}\epsilon\;,
\eea
for any two spinors $\chi, \epsilon$\,.
On top of the fields of maximal 5D supergravity, the ${\cal N}=(4,0)$ model also features exotic gravitino fields, or
fermionic two-forms $\psi_{\mu\nu}^A$\,. Their presence can be inferred from the decomposition of the 6D exotic gravitino fields
$\Psi_{\hat\mu\hat\nu}^A$, but also proves necessary for closure of the supersymmetry algebra in the 5+1 split.
In some sense (to be made precise in (\ref{deltaTC})), 
they play the role of the superpartners of the exotic tensor field ${\cal T}_{\mu\nu,\rho\sigma}$. 
In particular, after
reduction to 5D (i.e.\ in the limit $\partial_6\rightarrow0$), they have vanishing field strength
$\partial_{[\mu} \psi_{\nu\rho]}{}^A=0$, and can be set to zero by
fermionic gauge transformations
\bea
\delta \psi_{\mu\nu}^A &=& 2\,\partial_{[\mu} \kappa_{\nu]}{}^A
\;,
\label{ferm_gauge_5D}
\eea
in analogy to (\ref{gaugeT0}).

\subsection{${\cal N}=(4,0)$ supersymmetry}
\label{sec:n40susy}

Concretely, we impose the following supersymmetry transformation rules for the bosonic fields of the ${\cal N}=(4,0)$ model
\begin{align}
\delta e_{\mu}{}^\alpha ~=~&  \frac12\,\bar{\epsilon}_A \gamma^\alpha \psi_{\mu}^A \;, 
\nonumber\\
\delta \phi^{ABCD} ~=~& i\,\bar{\epsilon}^{\llbracket A} \chi^{BCD\rrbracket} \;, 
\nonumber\\
\delta A_\mu^{AB} ~=~&  \bar{\epsilon}_C \gamma_\mu \chi^{ABC} -i\, \bar{\epsilon}^{\llbracket A} \psi_\mu^{B \rrbracket}\;, 
\nonumber\\
\delta B_{\mu\nu}^{AB} ~=~& -i\, \bar{\epsilon}_C \gamma_{\mu\nu} \chi^{ABC}  +2\, \bar{\epsilon}^{\llbracket A}\gamma_{[\mu} \psi_{\nu]}^{B \rrbracket} -2 i\,\bar{\epsilon}^{\llbracket A} \psi_{\mu\nu}^{B \rrbracket} \;, 
\nonumber\\
\delta{{\cal C}}_{\mu\nu}{}^{\alpha} ~=~&
\bar{\epsilon}_A  \gamma^{\alpha} \psi_{\mu\nu}^A  +
i\,\bar{\epsilon}_A \left(2\, \eta_{\kappa[\mu} \gamma_{\nu]}{}^\alpha{}  - \eta_{\kappa[\mu} \delta_{\nu]}{}^{\alpha} 
\right) 
 \psi^{\kappa\,A}
\;,
 \nonumber\\
\delta {\cal T}_{\mu\nu}{}^{\alpha\beta} ~=~& 
 -2 i\, \bar{\epsilon}_A  \gamma^{\alpha\beta}\,  \psi_{\mu\nu}^A 
 \;,
 \label{susy_51_bos}
\end{align}
with a constant supersymmetry parameter $\epsilon^A$. On dimensional grounds, the supersymmetry transformation
of the bosonic fields is algebraic in the fermionic fields, in particular they remain unchanged in the $\partial_6\rightarrow0$ limit.
The first four lines of (\ref{susy_51_bos}) can thus be deduced from (linearized) maximal 5D supergravity~\cite{deWit:2004nw} ---
except for the last term in $\delta B_{\mu\nu}^{AB}$ which in the reduction to 5D becomes a gauge transformation (\ref{ferm_gauge_5D}).
The supersymmetry transformations of the exotic fields ${{\cal C}}_{\mu\nu}{}^{\alpha}$ and ${\cal T}_{\mu\nu}{}^{\alpha\beta}$ cannot be found in 5D supergravity and have to be derived by closure of the supersymmetry algebra.

To this end, we also need the fermionic supersymmetry transformation rules, which are given by
\begin{align}
\delta \chi^{ABC} ~=~& -i\,\gamma^\mu \epsilon_D \partial_\mu \phi^{ABCD}  
-\frac{3}{16}\, \gamma^{\mu\nu}\epsilon^{\llbracket A} {\cal F}_{\mu\nu}^{BC \rrbracket} 
- \partial_6 \phi^{ABCD} \epsilon_D 
\;,
\nonumber\\
\delta \psi_\mu^A ~=~& 
-\frac{i}{3}\, \gamma^\nu \epsilon_B {\cal F}_{\mu\nu}^{AB} 
+ \frac{i}{12}\, \gamma_\mu{}^{\sigma\tau} \epsilon_B {\cal F}_{\sigma\tau}^{AB} 
 - \frac{1}{16}\, \gamma^{\sigma\tau} \epsilon^A \left(
{\widehat\Omega_{\sigma\tau,\mu}}-2\,{\widehat\Omega_{\mu\sigma,\tau}}
\right)
\;,
\nonumber\\
\delta \psi_{\mu\nu}^A ~=~&  \frac{1}{2}  {\cal F}_{\mu\nu}^{AB} \epsilon_B 
 + \frac{i}{8}\, \gamma_\alpha \epsilon^A \,{\widehat\Omega_{\mu\nu}{}^{\alpha}} 
 -\frac{1}{32}\gamma_{\alpha\beta}\, \epsilon^A \partial_6 \left(  {{\cal T}}_{\mu\nu}{}^{\alpha\beta}
  -8\,\delta_{[\mu}{}^\alpha e_{\nu]}{}^{\beta} 
   \right)
\;.
\label{susy_51_ferm}
\end{align}
Again, most of the first two lines can be deduced from the supersymmetry transformation rules of maximal 5D supergravity
upon replacing the field strengths $F_{\mu\nu}{}^A$ and the 5D anholonomity objects $\Omega_{\mu\nu,\rho}$ by their
`covariantized' versions from (\ref{FB}) and (\ref{Omegahat_e}), respectively. The variation $\delta \psi_{\mu\nu}^A$ is
derived from closure of the algebra. In the $\partial_6\rightarrow0$ limit, this variation reduces to a gauge transformation
of type (\ref{ferm_gauge_5D}), consistent with the fact, that the exotic gravitino becomes pure gauge in 5D.

As a first consistency check of the supersymmetry transformation laws (\ref{susy_51_bos}), (\ref{susy_51_ferm}), we compute their
commutators on the bosonic fields which must close into the bosonic gauge transformations (\ref{deltaAB}), (\ref{gauge_wL}), (\ref{deltaT}),
together with global translations on all fields
\bea
\delta_{\rm transl} &=&
\Xi^\mu\partial_\mu 
+ \Xi^{6} \partial_6 
\;.
\eea
Specifically, in the commutator $[\delta_{\epsilon_1},\delta_{\epsilon_2}]$, we find closure with
global translation parameters
\bea
\Xi^\mu &=&
 \frac{1}{4}\, \bar{\epsilon}_{2,A} 
\gamma^{\mu} \epsilon_1^A
\;,\qquad
\Xi^6 ~=~
 -\frac{i}{4} \, \bar{\epsilon}_{2,A} 
 \epsilon_1^A
\;,
\eea
and local gauge parameters 
\bea
\xi_\mu &=&
- \Xi^\rho e_{\rho\mu}
\;,
\qquad
\Gamma_{\mu\nu}~=~
-\Xi^{\rho}  \,{{\cal C}}_{\rho\mu,\nu}
+\Xi^6  e_{\mu\nu}
\;,
\\
  \Lambda_{\rho,\mu\nu} &=& 
   -  \Xi^{\sigma}
{{\cal T}}_{\sigma\rho,\mu\nu} 
-4\, e_{\rho[\mu} \Xi_{\nu]}
\;,
\qquad
\Lambda^{AB}~=~
-2\,i\,\bar{\epsilon}_{C,2}  \epsilon_{D,1}  \phi^{ABCD} 
\;,
\nonumber\\
\Lambda_{\mu\nu} &=&
 -\frac{1}{2}\, \Xi^\tau\left( 
 \widehat{\Omega}_{\mu\nu,\tau} 
- 2\, \widehat{\Omega}_{\tau[\mu,\nu]} 
  \right) 
+\frac{i}{3}\, \bar{\epsilon}_{2,A}  \epsilon_{1,B} {\cal F}_{\mu\nu}^{AB} 
+ \frac{1}{12}\, \varepsilon_{\mu\nu}{}^{\sigma\tau\lambda} \bar{\epsilon}_{2,A} \gamma_{\lambda} \epsilon_{1,B} {\cal F}_{\sigma\tau}^{AB} 
\;.
\nonumber
\eea
It is worth to note that closure on the bosonic fields
$B_{\mu\nu}^{AB}$, ${\cal C}_{\mu\nu}{}^{\alpha}$, and ${\cal T}_{\mu\nu}{}^{\alpha\beta}$,
requires the use of their field equations (\ref{eomFH}), (\ref{eomCC}), and (\ref{eomTT}),
respectively. This is to be expected (and familiar from supergravity), as these fields obey first order duality equations.

\subsection{Supersymmetric Lagrangian}

Equipped, with the supersymmetry transformation laws (\ref{susy_51_bos}), (\ref{susy_51_ferm}), 
we can now present the full ${\cal N}=(4,0)$ supersymmetric Lagrangian, extending the bosonic result 
(\ref{ExFTsd}), (\ref{frameinvLag}). The final result is given by
\bea
{\cal L} &=&
-\frac{1}{4}\,\widehat{\Omega}^{\mu\nu\rho} \widehat{\Omega}_{\mu\nu\rho}
+\frac12\,\widehat{\Omega}^{\mu\nu\rho}\widehat{\Omega}_{\nu\rho\mu}
+\widehat{\Omega}^{\mu}\widehat{\Omega}_{\mu}
- \frac{8}{3}\,\partial^\mu \phi_{ABCD} \partial_\mu \phi^{ABCD}
-\frac{1}{2}\, {\cal F}^{\mu\nu}{}_{AB} {\cal F}_{\mu\nu}{}^{AB} 
\nonumber\\
&&{}
 - \frac14\,
\varepsilon^{\mu\nu\rho\sigma\tau} 
\partial_\mu B_{\nu\rho,AB}
\partial_6 B_{\sigma\tau}{}^{AB} 
- \frac1{8}\,\varepsilon^{\mu\nu\rho\sigma\tau}\,
\partial_{\mu} {{\cal C}}{}_{\nu\rho,\alpha} \,\partial_6  {{\cal C}}{}_{\sigma\tau}{}^\alpha
-\frac1{64}\,
\varepsilon^{\mu\nu\rho\sigma\tau} 
\partial_\mu {\cal T}_{\nu\rho,\alpha\beta}
\partial_6 {\cal T}_{\sigma\tau}{}^{\alpha\beta} 
\nonumber\\
&&{}
-\frac83\,\partial_6 \phi_{ABCD} \partial_6 \phi^{ABCD}
  +3\,
\partial_6 e_{\mu}{}^{\sigma}
\, \partial_6 e_{\sigma}{}^{\mu} 
-3\,
\partial_6 e_{\mu}{}^{\mu}
\, \partial_6 e_\nu{}^\nu
+
\frac1{96}\,
\varepsilon^{\mu\nu\rho\sigma\tau} 
\partial_6 {\cal T}_{\mu\nu,\alpha\beta}
 \partial_6 {\cal J}'_{\rho\sigma\tau}{}^{\alpha\beta} 
\nonumber\\
 &&{}
 + \frac{1}{2} \,\varepsilon^{\mu\nu\rho\alpha\beta} \partial_6 e_{\alpha\beta} \partial_{\mu} e_{\nu\rho}
  +\frac{1}{2}\, \varepsilon^{\mu\nu\rho\alpha\beta} \partial_6 e_{\alpha\beta} \partial_6 {{\cal C}}_{\mu\nu,\rho}
\nonumber\\
&&{}
-\frac83\,\bar\chi_{ABC}  \gamma^{\mu} \partial_\mu \chi^{ABC}
-2\,\bar\psi_{\mu\,A}  \gamma^{\mu\nu\rho} \partial_\nu \psi_{\rho}{}^{A}
+ \bar\psi_{\mu\nu\,A} \gamma^{\mu\nu\rho\sigma\tau} \partial_\rho \psi_{\sigma\tau}{}^{A}
\nonumber\\
&&{}
-\frac{8 i}{3}\, \bar\chi_{ABC}  \, \partial_6 \chi^{ABC} + 6\,i\,\bar\psi_{\mu\,A}  \gamma^{\mu\nu} \partial_6 \psi_{\nu}^{A} 
+i\, \bar\psi_{\mu\nu\,A} \gamma^{\mu\nu\rho\sigma} \partial_6 \psi_{\rho\sigma}^{A}+
 4\, \bar\psi_{\mu\nu\,A}  \gamma^{\mu\nu\rho} \partial_6 \psi_{\rho}^{A}
 \;,
\label{L40}
\eea
where ${\cal F}_{\mu\nu}^{AB}$, $\widehat{\Omega}_{\mu\nu\rho}$, and ${\cal J}'_{\rho\sigma\tau}{}^{\alpha\beta}$ have been defined
in (\ref{FB}), and (\ref{Omegahat_e}), respectively. This Lagrangian can be shown to be invariant under the supersymmetry
transformations (\ref{susy_51_bos}), (\ref{susy_51_ferm}), up to total derivatives. A number of comments are in order:
\begin{itemize}
\item
The first line of (\ref{L40}) is the linearized version of the bosonic sector of 5D maximal supergravity with the different objects
`covariantized' by $\partial_6$ contributions accoding to (\ref{FB}) and (\ref{Omegahat_e}).
Similarly, the first two terms of the fifth line of (\ref{L40}) represent the linearized 
version of the fermionic sector of 5D maximal supergravity.

\item
Contrary to the bosonic fields $B_{\mu\nu}^{AB}$, ${\cal C}_{\mu\nu}{}^{\alpha}$, and ${\cal T}_{\mu\nu}{}^{\alpha\beta}$
which drop out of (\ref{L40}) in the limit $\partial_6\rightarrow0$, the exotic gravitino field $\psi_{\mu\nu}^A$ survives this limit
with a non-trivial Rarita-Schwinger type kinetic term
$\bar\psi_{\mu\nu\,A} \gamma^{\mu\nu\rho\sigma\tau} \partial_\rho \psi_{\sigma\tau}{}^{A}$.
However, its field equations in this limit imply its vanishing curvature $\partial_{[\mu} \psi_{\nu\rho]}{}^A=0$, such 
that it can be set to zero by fermionic gauge transformations (\ref{ferm_gauge_5D}) as anticipated above.

\item
From the 5D perspective, the second line of (\ref{L40}) represents a number of topological Chern-Simons terms for two forms
$\{B_{\mu\nu}, {\cal C}_{\mu\nu}, {\cal T}_{\mu\nu}\}$, with the antisymmetric metric given by 
$\langle X,Y \rangle \propto \int dy\,X\partial_6 Y$, together with index contraction over Lorentz and USp(8) indices.
The form of these terms may be read as a hint toward introducing interactions via putative non-linear gauge structures.

\item
Up to total derivatives, the Lagrangian (\ref{L40}) is also 
invariant under gauge transformations with fermionic gauge parameters
\bea
\delta \psi_{\mu\nu}^A &=& 
2\,\partial_{[\mu} \kappa_{\nu]}^A
+ i\,\partial_{[\mu} \gamma_{\nu]} \kappa^A
+2\,i\,\partial_6 \gamma_{[\mu} \kappa^A_{\nu]}
\;,\nonumber\\
\delta \psi_\mu^A &=& \partial_\mu \kappa^A-2\,\partial_6 \kappa_\mu^A 
\;,
\label{ferm_gauge}
\eea
which extend (\ref{ferm_gauge_5D}) in case of non-trivial $\partial_6$\,.
In an interacting theory, the fermionic gauge parameter $\kappa^A$ would be expected
to fuse with the parameter $\epsilon^A$ of global supersymmetry, just as in
standard supergravity.

\end{itemize}

\section{Embedding into 6D}

In the last section, we have constructed the ${\cal N}=(4,0)$ supersymmetric extension of the bosonic Lagrangian of the model
in the 5+1 split of coordinates, with all couplings uniquely determined by closure of the supersymmetry algebra.
In this section, we will show that the fermionic equations of motion derived from the Lagrangian (\ref{L40}),
as well as the supersymmetry transformations (\ref{susy_51_bos}), (\ref{susy_51_ferm}) indeed lift up to the original six-dimensional
field equations and supersymmetry transformation rules.
To this end, we first work out the 6D supersymmetry transformation rules, decompose them according to the 5+1 
coordinate split and modify them by on-shell vanishing contributions and fermionic gauge transformations
in order to establish the equivalence with our previous results.
Appendix~\ref{app:6D} summarizes our 6D spinor conventions.

\subsection{The 6D model and ${\cal N}=(4,0)$ supersymmetry variations}

The 6D ${\cal N}=(4,0)$ supermultiplet combines the following field content \cite{Strathdee:1986jr}
\bea
\begin{array}{cc}
    \toprule
    \text{field} & G     \\
    \midrule
      \phi^{ABCD}  & (1,1;42)   \\
      \chi^{ABC}  &  (2,1;48) \\
     B_{\hat\mu\hat\nu}^{AB} &  (3,1;27)   \\
      \psi_{\hat\mu\hat\nu}^A &  (4,1;8) \\
      T_{\hat\mu\hat\nu,\hat\rho\hat\sigma}  &  (5,1;1)  \\
    \bottomrule 
\end{array}
\label{sm6D}
\eea
where the second column denotes the representation of the various fields under 
the product of little group and R-symmetry $G = \mathrm{SU(2)\times SU(2) \times USp(8)}$.
All fields carrying multiple USp(8) indices $A, B, \dots$, are totally antisymmetric and symplectic traceless in these
indices.
The bosonic field equations have been discussed in section~\ref{subsec:split40} above and comprise the self-duality equations
\bea
H_{\hat\mu\hat\nu\hat\rho}{}^{AB} &=& 
\frac16\,\varepsilon_{\hat\mu\hat\nu\hat\rho\hat\sigma\hat\kappa\hat\lambda}\,H^{\hat\sigma\hat\kappa\hat\lambda}{}^{AB}
\;,
\nonumber\\
G_{\hat{\mu}\hat{\nu}\hat{\lambda},\hat{\rho}\hat{\sigma}\hat{\tau}} &=& 
\frac16\,\varepsilon_{\hat{\mu}\hat{\nu}\hat{\lambda}\hat{\alpha}\hat{\beta}\hat{\gamma}}\,
G^{\hat{\alpha}\hat{\beta}\hat{\gamma}}{}_{\hat{\rho}\hat{\sigma}\hat{\tau}}\;,
\label{6D_bos}
\eea
for the first order field strength $H_{\hat\mu\hat\nu\hat\rho}{}^{AB}=3\,\partial_{[\hat\mu}B_{\hat\nu\hat\rho]}{}^{AB}$
and the second order field strength (\ref{defGT}), respectively. The fermionic field equations combine a standard Dirac equation
for the spin-1/2 fermions $\chi^{ABC}$, together with a Rarita-Schwinger like equation 
\bea
 \hat{\gamma}^{\hat{\mu}\hat{\nu}\hat{\rho}\hat{\sigma}\hat{\tau}} \partial_{\hat{\rho}}\Psi_{\hat{\sigma}\hat{\tau}}^A =0
\;,
\label{p26}
\eea
for the {exotic gravitini} or fermionic two-forms $\psi_{\hat\mu\hat\nu}^A$, c.f.~\cite{Buchbinder:2009pa,Zinoviev:2009wh,Lekeu:2021oti},
invariant under gauge transformations
\bea
\delta \Psi_{\hat{\mu}\hat{\nu}}&=& 2\,\partial_{[\hat\mu} \kappa_{\hat\nu]}
\;.
\label{6D_gauge_ferm} 
\eea
All fermionic fields are of positive chirality in 6D, whereas the supersymmetry parameter $\epsilon^A$ carries negative chirality.

While the supermultiplet (\ref{sm6D}) was constructed in \cite{Strathdee:1986jr}, to the best of our knowledge the explicit 6D supersymmetry transformation laws have not been spelled out in components before\footnote{
They have however been constructed in the pre-potential formalism in~\cite{Henneaux:2017xsb}, see also
\cite{Cederwall:2020dui} for  a superspace construction.}.
We construct them here, by imposing closure of the supersymmetry algebra and find
\begin{align}
\delta \phi^{ABCD} &=i\, \bar{\epsilon}^{\llbracket A} \chi^{BCD \rrbracket} 
\;,\nonumber\\
\delta \chi^{ABC} &= -i\,\hat{\gamma}^{\hat{\mu}} \partial_{\hat{\mu}} \phi^{ABCD}\epsilon_D 
-\frac{i}{32}\, \hat{\gamma}^{\hat{\mu}\hat{\nu}\hat{\rho}}H_{\hat{\mu}\hat{\nu}\hat{\rho}}{}^{\llbracket AB} \epsilon^{C \rrbracket} 
\;,\nonumber\\
\delta B_{\hat{\mu}\hat{\nu}}^{AB} &= 
-i\, \bar{\epsilon}_C \hat{\gamma}_{\hat{\mu}\hat{\nu}} \chi^{ABC} 
- 2\,i\,\bar{\epsilon}^{\llbracket A} \Psi_{\hat{\mu}\hat{\nu}}^{B \rrbracket} 
\;,\nonumber\\
\delta \Psi_{\hat{\mu}\hat{\nu}}^A &= 
\frac{i}{6}\, \hat{\gamma}^{\hat{\rho}} H_{\hat{\mu}\hat{\nu}\hat{\rho}}{}^{AB} \epsilon_B 
-\frac{i}{12}\, H_{ \hat{\alpha}\hat{\beta}[\hat{\mu}}{}^{AB}\hat{\gamma}_{\hat{\nu}]}{}^{\hat{\alpha}\hat{\beta}} \epsilon_B 
-\frac{i}{192}\,\hat{\gamma}^{\hat{\alpha}\hat{\beta}\hat{\gamma}}S_{\hat{\alpha}\hat{\beta}\hat{\gamma},\hat{\mu}\hat{\nu}}\epsilon^A 
\;,\nonumber\\
\delta T_{\hat{\mu}\hat{\nu},\hat{\rho}\hat{\sigma}} &= 
-i\,\bar{\epsilon}_A \left(\hat{\gamma}_{\hat{\mu}\hat{\nu}}\psi_{\hat{\rho}\hat{\sigma}}^A + \hat{\gamma}_{\hat{\rho}\hat{\sigma}}\psi_{\hat{\mu}\hat{\nu}}^A - 2\,\hat{\gamma}_{[\hat{\mu}\hat{\nu}}\psi_{\hat{\rho}\hat{\sigma}]}^A \right)
\;,
\label{6D_susy}
\end{align}
unique on-shell up to rescaling. Here, $S_{\hat{\alpha}\hat{\beta}\hat{\gamma},\hat{\mu}\hat{\nu}}$ is the first order field strength
\begin{equation}
S_{\hat{\mu}\hat{\nu}\hat{\rho},\hat{\sigma}\hat{\tau}} = 3\, \partial_{[\hat\mu} T_{\hat{\nu}\hat{\rho}],\hat{\sigma}\hat{\tau}} \;,
\end{equation}
which is not gauge invariant under the transformations (\ref{gauge40}), however its variation induces a fermionic
gauge transformation (\ref{6D_gauge_ferm}), justifying its appearance on the r.h.s.\ of (\ref{6D_susy}).
The different terms in the variation of $T_{\hat{\mu}\hat{\nu},\hat{\rho}\hat{\sigma}}$ precisely implement the ${\tiny \yng(2,2)}$
Young tableau symmetry of the field.

The algebra of supersymmetry transformations (\ref{6D_susy}) closes on scalar fields according to
\bea
[\delta_1, \delta_2 ] \,\phi^{ABCD} &=&
\bar{\epsilon}_2^{\llbracket A} \hat{\gamma}^{\hat{\mu}}\epsilon_{1,E} \partial_{\hat{\mu}} \phi^{BCD\rrbracket E} - \bar{\epsilon}_1^{\llbracket A} \hat{\gamma}^{\hat{\mu}}\epsilon_{2,E} \partial_{\hat{\mu}} \phi^{BCD\rrbracket E}  
\nonumber\\
&=& -\frac{1}{4}\,\bar{\epsilon}_2^{E} \hat{\gamma}^{\hat{\mu}}\epsilon_{1,E} \partial_{\hat{\mu}} \phi^{ABCD}
\;,
\eea
after using a Schouten identity on the r.h.s. This is the action of a diffeomorphism with parameter
\begin{equation}
\xi^{\hat{\mu}} = -\frac{1}{4}\, \bar{\epsilon}_2^{E} \hat{\gamma}^{\hat{\mu}}\epsilon_{1,E}
\;.
\end{equation}
Similarly, one may verify closure on the other fields, where use of the first order equations of motion 
(\ref{6D_bos}), (\ref{p26}), are required for closure on the respective fields.
Since there is no standard 6D covariant action principle for the bosonic field equations (\ref{6D_bos}),
all supersymmetry transformations (\ref{6D_susy}) are only defined on-shell.

Furthermore, the algebra only closes up to gauge transformations, \emph{i.e.} for the commutator of the $B_{\hat{\mu}\hat{\nu}}^{AB}$-field we find,
\begin{align}
[\delta_1,\delta_2] B_{\hat{\mu}\hat{\nu}}^{AB} = \xi^{\hat{\rho}} H_{\hat{\mu}\hat{\nu}\hat{\rho}}^{AB} -4\,\bar{\epsilon}_{2,C} \,\hat{\gamma}_{[\hat{\mu}}\,\epsilon_{D,1}\,\partial_{\hat{\nu}]}\,\phi^{ABCD}\,,
\end{align}
where the second term is a six-dimensional gauge variation.


\subsection{5+1 split of the fermionic equations of motion}

In this section, we work out the 6D fermionic equations of motion according to the 5+1 split
of coordinates (\ref{split51}) in analogy to the discussion of the bosonic equations of motion in section~\ref{subsec:split40} above.
For the gamma matrices, we take the conventions from \cite{Henneaux:2017xsb}, see also appendix~\ref{app:6D}.
The 5D gamma matrices are chosen as set of five $4\times 4$ matrices
\begin{equation}
(\gamma_0,\dots ,\gamma_3, \gamma_4 = i\, \gamma_{0123})
\;.
\label{gamma5D}
\end{equation}
The first four matrices are a representation of the 4D Clifford algebra ${\cal C}(1,3)$ and $\gamma_4$ is the chirality matrix associated to them. 
The 6D gamma matrices, which we denote by $\hat\gamma_{\hat\mu}$, can be built from (\ref{gamma5D}) 
by defining the following family of six $8\times 8$ matrices
\begin{align}
\label{gamma6}
\hat{\gamma}_\mu &= \sigma_1 \otimes \gamma_\mu = \begin{pmatrix}
0 & \gamma_\mu \\
\gamma_\mu & 0
\end{pmatrix}, \qquad
\hat{\gamma}_6 = \sigma_2 \otimes \mathbb{I} = \begin{pmatrix}
0 & -i \,\mathbb{I} \\
i \, \mathbb{I} & 0
\end{pmatrix} \;,\\ 
& \Longrightarrow\;\;\;\hat{\gamma}_7 = \hat{\gamma}_{012346} = \begin{pmatrix}
\mathbb{I} & 0 \\
0 & -\mathbb{I}
\end{pmatrix}.
\end{align}
All 6D fermions ${\cal X}\in\{\Psi_{\hat\mu\hat\nu}^A, \chi^{ABC}\}$ 
are of positive chirality, whereas the supersymmetry parameter $\epsilon$ is of negative chirality,
i.e.\ in this basis they are of the block form (with slight abuse of notation)
\begin{equation}
{\cal X} = \begin{pmatrix}
{\cal X} \\
0
\end{pmatrix}
\;,\qquad
\epsilon = \begin{pmatrix}
 0\\
\epsilon
\end{pmatrix}
\;.
\label{chiralF}
\end{equation}

In this basis, the 6D Dirac equation of the spin-1/2 fermions $\chi^{ABC}$ takes the form
\begin{equation}
\hat{\gamma}^{\hat{\mu}} \partial_{\hat{\mu}} \chi^{ABC} \quad \Longleftrightarrow \quad \gamma^\mu \partial_\mu \chi^{ABC} = - i\,\partial_6 \chi^{ABC}\;,
\end{equation}
which is precisely the equation following from variation of the Lagrangian (\ref{L40}).
For the exotic gravitini $\Psi_{\hat{\mu}\hat{\nu}}^{A}$ their field equation (\ref{p26}) splits according to
\begin{equation}
 \begin{cases}
 &\gamma^{\mu\nu\rho\sigma\tau} \partial_\rho (\Psi_{\sigma\tau}^{A} + 2i\,\gamma_\tau  \Psi_{\sigma 6}^{A}) = - i\, \partial_6 \gamma^{\mu\nu\rho\sigma} \Psi_{\rho\sigma}^{A}\;, \\
&\gamma^{\mu\nu\rho\sigma} \partial_\nu \Psi_{\rho\sigma}^{A} = 0 \;.
\end{cases}
\label{eomg22}
\end{equation}
Defining
\begin{align}
\psi_\mu^{A} =  2\,\Psi_{\mu 6}^{A}\;,\qquad
\psi_{\mu\nu}^{A} = \Psi_{\mu\nu}^{A} - 2i\, \gamma_{[\mu}^{\vphantom{a}}\Psi_{\nu]6}^{A}\;,
\label{psipsi}
\end{align}
equations (\ref{eomg22}) become
\begin{align}
\gamma^{\mu\nu\rho\sigma\tau} \partial_\rho \psi_{\sigma\tau}^{A} =\;& 
-\partial_6 \left( i\, \gamma^{\mu\nu\rho\sigma}\psi_{\rho\sigma}^{A} + 2\, \gamma^{\mu\nu\rho} \psi_\rho^{A} \right)\;, \nonumber\\
\gamma^{\mu\nu\rho}\partial_\nu \psi_\rho^{A} =\;&  \partial_6 \left( 3i\, \gamma^{\mu\nu} \psi_\nu^{A} -\gamma^{\mu\nu\rho} \psi_{\nu\rho}^{A}  \right)\;,
\end{align}
and reproduce the corresponding field equations obtained from the Lagrangian (\ref{L40}).
Finally, the 6D fermionic gauge symmetry (\ref{6D_gauge_ferm}) associated with the exotic gravitini,
under the 5+1 split and with (\ref{psipsi}) translates into the gauge transformations
\bea
\delta \psi_{\mu\nu}^A &=& 
2\,\partial_{[\mu} \kappa_{\nu]}^A
+ i\,\partial_{[\mu} \gamma_{\nu]} \kappa^A
+2\,i\,\partial_6 \gamma_{[\mu} \kappa^A_{\nu]}
\;,\nonumber\\
\delta \psi_\mu^A &=& \partial_\mu \kappa^A-2\,\partial_6 \kappa_\mu^A 
\;,
\label{ferm_gauge_1}
\eea
with fermionic gauge parameters 
$\kappa^A_{\hat\mu}=\left\{\kappa^A_\mu,\frac12\kappa^A\right\}$\,. This precisely reproduces (\ref{ferm_gauge}),
identified above.
The field equations and gauge transformations of the Lagrangian (\ref{L40}) thus reproduce
the full ${\cal N}=(4,0)$ model in six dimensions.

\subsection{5+1 split of the supersymmetry transformations}

In this section, we also decompose the 6D supersymmetry transformations (\ref{6D_susy}) according to the 5+1
split and match the resulting structures to the transformation laws presented in section~\ref{sec:n40susy} above.
To accomplish the match, we need to modify the transformation laws obtained from (\ref{6D_susy})
by contributions proportional to the first order bosonic field equations as well as by fermionic gauge transformations 
(\ref{ferm_gauge_1}). This appears judicious, as the transformations (\ref{6D_susy}) are defined on-shell only
(leaving invariant a set of equations of motion rather than a Lagrangian), whereas the transformation rules
(\ref{susy_51_bos}), (\ref{susy_51_ferm}) define the symmetries of a Lagrangian (\ref{L40}).

In a first step, we evaluate the transformation rules (\ref{6D_susy}) for the different components 
of the bosonic fields as embedded within the 6D fields (\ref{AB6}), (\ref{TCh40}). The result is given by
\begin{align}
\delta \phi^{ABCD} &= i\,\bar{\epsilon}^{\llbracket A} \chi^{BCD\rrbracket} 
\;,\nonumber\\
\delta A_\mu^{AB} &=  \bar{\epsilon}_C \gamma_\mu \chi^{ABC} -i\, \bar{\epsilon}^{\llbracket A} \psi_\mu^{B \rrbracket} 
\;,\nonumber\\
\delta B_{\mu\nu}^{AB} &= -i\, \bar{\epsilon}_C \gamma_{\mu\nu} \chi^{ABC}  +2\, \bar{\epsilon}^{\llbracket A}\gamma_{[\mu} \psi_{\nu]}^{B \rrbracket} -2 i\,\bar{\epsilon}^{\llbracket A} \psi_{\mu\nu}^{B \rrbracket} 
\;,\nonumber\\
\delta h_{\mu\nu} &=  \bar{\epsilon}_A \gamma_{(\mu} \psi_{\nu)}^A 
\;,\nonumber\\
\delta C_{\mu\nu,\rho} &= \bar{\epsilon}_A \left( ( \gamma_\rho \psi_{\mu\nu}^A - \gamma_{[\rho} \psi_{\mu\nu]}^A) -i\, (\gamma_{\mu\nu}\psi_\rho^A - \gamma_{[\mu\nu}\psi_{\rho]}^A ) -i\, \psi_{[\mu}^A \eta_{\nu]\rho} \right) 
\;,\nonumber\\
\delta T_{\mu\nu,\rho\sigma} &= 
-i\, \bar{\epsilon}_A \left( \gamma_{\mu\nu} \psi_{\rho\sigma}^A + \gamma_{\rho\sigma} \psi_{\mu\nu}^A -  2\,\gamma_{[\mu\nu} \psi_{\rho\sigma]}^A \right) 
+2\, \bar{\epsilon}_A \left( \gamma_{[\mu} \eta_{\nu][\rho} \psi_{\sigma]}^A + \gamma_{[\rho} \eta_{\sigma][\mu} \psi_{\nu]}^A \right) 
\;,
\label{susy_bos_65}
\end{align}
with the first four lines already reproducing the corresponding transformations in (\ref{susy_51_bos}) above.
In order to also reproduce the previous results for ${\cal C}_{\mu\nu}{}^\alpha$ and ${\cal T}_{\mu\nu,\alpha\beta}$,
we first need to determine the transformation laws for the fields $u_{\mu\nu}$ and $v_{\rho\mu\nu}$,
defined by the first order equations (\ref{defv}) and (\ref{defu40}), respectively. As a consistency check of the whole construction,
one can successively compute the variation of the l.h.s.\ of these equations according to (\ref{susy_bos_65}) and finds after some
lengthy computation and using the fermionic field equations, that its variation is indeed given by a curl in 
$[\sigma\tau]$ for (\ref{defv}) and by a gradient in $\rho$ for (\ref{defu40}).
Upon integration, this eventually yields the supersymmetry transformations
\bea
\delta v_{\tau,\mu\nu} &=&
 \bar{\epsilon}_A  \gamma_{\mu\nu}{}^{\rho}\,  \psi_{\tau\rho}^A 
+ \bar{\epsilon}_A \gamma_{\tau}  \psi_{\mu\nu}^A 
-2 \, \bar{\epsilon}_A \gamma_{[\tau}   \psi_{\mu\nu]}^A 
   +\frac13\,\eta_{\tau[\mu}\, \bar{\epsilon}_A  \gamma_{\nu]}{}^{\alpha\beta}\, \psi_{\alpha\beta}^{A} 
   \nonumber\\
   &&{}
 -\frac12\,i\, \bar{\epsilon}_A \gamma_{\mu\nu\tau\gamma}  \psi^{\gamma\,A} 
-i\,  \bar{\epsilon}_A\gamma_{\mu\nu}  \psi_{\tau}^A
-i\, \bar{\epsilon}_A \gamma_{[\mu\nu}  \psi_{\tau]}^A 
+i\,  \eta_{\tau[\mu}   \bar{\epsilon}_A \psi_{\nu]}^A 
\;,\nonumber\\[1ex]
\delta u_{\mu\nu} &=&
- \frac12\,  \bar\epsilon_A \gamma_{[\mu} \psi_{\nu]}^A
 -\frac13  \bar{\epsilon}_A \gamma_{\mu\nu}{}^{\rho}   \psi_\rho^A 
 - \frac1{12} i \, \bar{\epsilon}_A \gamma_{\mu\nu}{}^{\rho\sigma}    \psi_{\rho\sigma}^A 
\;.
\eea
From this we obtain the variation of the fields (\ref{newT}) and (\ref{newC}) as
\bea
\delta {\cal T}_{\mu\nu,\rho\sigma} &=& 
 -2 i\, \bar{\epsilon}_A  \gamma_{\rho\sigma}\,  \psi_{\mu\nu}^A 
\;,
\nonumber\\
\delta \widehat{C}_{\mu\nu,\rho} &=&
\bar{\epsilon}_A  \gamma_\rho \psi_{\mu\nu}^A  +
\frac{i}{2}\,\bar{\epsilon}_A \left(4\, \gamma_{\rho[\mu}\eta_{\nu]\kappa}  -2\, \eta_{\kappa[\mu} \eta_{\nu]\rho} 
+ \gamma_{\mu\nu\rho\kappa}\right) 
 \psi^{\kappa\,A}\;,
 \label{deltaTC}
\eea
in agreement with (\ref{susy_51_bos}) above.
In order to compare the variation of $\widehat{C}_{\mu\nu,\rho}$, we should fix the Lorentz gauge freedom  in (\ref{susy_51_bos})
by imposing $e_{[\mu\nu]}=0$, upon which we can identify $\widehat{C}_{\mu\nu,\rho}$ and 
${\cal C}_{\mu\nu,\rho}$ according to (\ref{e00}), and the variation of the latter acquires an extra term from a
compensating Lorentz transformation. The final expressions then coincide.
Let us stress the simplicity of the supersymmetry transformation of the reducible field ${\cal T}_{\mu\nu,\rho\sigma}$  in (\ref{deltaTC})
compared to the transformation (\ref{susy_bos_65}) of the irreducible field ${T}_{\mu\nu,\rho\sigma}$ embedded in 6D.

Next, we can work out the fermionic transformation rules. After evaluating (\ref{6D_susy}) for the different components 
of the fermionic fields as embedded within the 6D fields (\ref{psipsi}), we obtain
\begin{align}
\delta \chi^{ABC} &= -i\,\gamma^\mu \epsilon_D \partial_\mu \phi^{ABCD} - \partial_6 \phi^{ABCD} \epsilon_D -\frac{3}{16}\, \gamma^{\mu\nu}\epsilon^{\llbracket A} (F_{\mu\nu}^{BC \rrbracket} + \partial_6 B_{\mu\nu}^{BC \rrbracket})  
\;,\nonumber\\
\delta \psi_\mu^A &= -\frac{i}{3}\, \gamma^\nu \epsilon_B (F_{\mu\nu}^{AB} + \partial_6 B_{\mu\nu}^{AB}) + \frac{i}{6}\, \gamma_\mu{}^{\alpha\beta} \epsilon_B (F_{\alpha\beta}^{AB} + \partial_6 B_{\alpha\beta}^{AB}) \nonumber \\
&\quad - \frac{1}{8}\, \gamma^{\alpha\beta} \epsilon^A \left(\partial_\alpha h_{\beta\mu} + \frac{1}{2}\partial_6 C_{\alpha\beta,\mu} + \frac{1}{4} \partial_6 v_{\mu,\alpha\beta} - \frac{1}{2}\partial_\mu u_{\alpha\beta} \right) 
\;,\nonumber\\
\delta \psi_{\mu\nu}^A &=  \frac{1}{2}  (F_{\mu\nu}^{AB} + \partial_6 B_{\mu\nu}^{AB})\epsilon_B 
+ \frac{i}{8}\, \gamma^\alpha \epsilon^A \left( \partial_{[\mu} h_{\nu]\alpha} - \partial_6 C_{\alpha[\mu,\nu]} + \frac{1}{2}\partial_6 v_{[\nu,\mu]\alpha} + \partial_{[\mu} u_{\nu]\alpha} \right) \nonumber \\
&\quad  - \frac{1}{16}\, \gamma^{\alpha\beta}\epsilon^A \partial_{[\mu} (C_{|\alpha\beta|,\nu]} - \frac{1}{2} v_{\nu],\alpha\beta}) -\frac{i}{16}\, \gamma_{[\nu}{}^{\alpha\beta}\epsilon^A \partial_{\mu]}u_{\alpha\beta} \nonumber \\
&\quad -\frac{1}{32}\gamma^{\alpha\beta}\epsilon^A \partial_6 T_{\alpha\beta,\mu\nu} - \frac{i}{32}\gamma_{[\mu}{}^{\alpha\beta}\epsilon^A \partial_6 v_{\nu],\alpha\beta}
\;,
\end{align}
where we have furthermore used the bosonic first order equations (\ref{eomFH}), (\ref{defv}), (\ref{defu40})
in order to bring the r.h.s.\ closer to the desired form of (\ref{susy_51_ferm}). The result can further be
simplified after combining these transformations with fermionic gauge transformations (\ref{ferm_gauge_1}) with parameters
\begin{align}
\kappa^A &= \frac{1}{16}\, \gamma^{\mu\nu} \epsilon^A u_{\mu\nu}, \nonumber\\
\kappa_\mu^A &= -\frac{1}{32}\gamma^{\sigma\tau} \epsilon^A \left( C_{\sigma\tau,\mu} - \frac{1}{2}\, v_{\mu,\sigma\tau} 
- \varepsilon_{\mu\kappa\lambda\sigma\tau} u^{\kappa\lambda} \right),
\end{align}
upon which we find
\begin{align}
\delta \psi_\mu^A &= -\frac{i}{3}\, \gamma^\nu \epsilon_B (F_{\mu\nu}^{AB} + \partial_6 B_{\mu\nu}^{AB}) + \frac{i}{6}\, \gamma_\mu{}^{\alpha\beta} \epsilon_B (F_{\alpha\beta}^{AB} + \partial_6 B_{\alpha\beta}^{AB}) \nonumber \\
&\quad - \frac{1}{8}\, \gamma^{\alpha\beta} \epsilon^A \left(\partial_\alpha h_{\beta\mu} +\partial_6 \left( C_{\alpha\beta,\mu} - \frac{1}{2}\, \varepsilon_{\mu\alpha\beta\kappa\lambda} u^{\kappa\lambda} \right) \right), \nonumber\\[2ex]
\delta \psi_{\mu\nu}^A &=  \frac{1}{2}  (F_{\mu\nu}^{AB} + \partial_6 B_{\mu\nu}^{AB})\epsilon_B
 + \frac{i}{8}\, \gamma^\alpha \epsilon^A \left( \partial_{[\mu} h_{\nu]\alpha} + \partial_6 \left( C_{\mu\nu,\alpha} + \varepsilon_{\mu\nu\alpha\kappa\lambda}u^{\kappa\lambda} \right) \right) \nonumber \\
&\quad -\frac{1}{32}\gamma^{\alpha\beta}\epsilon^A \partial_6 \left( T_{\alpha\beta,\mu\nu} + \varepsilon_{\alpha\beta\kappa\lambda [\mu} \left( C^{\kappa\lambda}{}_{\nu]} - v_{\nu],}{}^{\kappa\lambda} - \varepsilon^{\kappa\lambda}{}_{\nu]\rho\sigma} u^{\rho\sigma} \right) \right),
\end{align}
which precisely reproduces (\ref{susy_51_ferm}) after recombining the fields on the r.h.s.\
according to the definitions~(\ref{newT}), (\ref{newC}), (\ref{Omegahat}).

We have thus shown that the supersymmetry transformation rules (\ref{susy_51_bos}), (\ref{susy_51_ferm}),
which leave the Lagrangian (\ref{L40}) invariant, indeed descend from the supersymmetry transformation laws
in six dimensions.

\section{Conclusions}

In this paper we have given the complete free action for $(4,0)$ supergravity in 6D, using a $5+1$ split of coordinates. 
This includes the fermionic fields such as  exotic gravitini, as well as the supersymmetry rules. 
Due to the $5+1$ split, the 6D Lorentz invariance is not manifest, but the theory is fully equivalent to 
$(4,0)$ supergravity at the free level, as follows by integrating the equations of motion following from this action. 
For the special case that the sixth coordinate parameterizes a circle, this action encodes the dynamics of 
all Kaluza-Klein modes that in turn organize into BPS multiplets of the supersymmetry algebra~\cite{Hull:2000cf}.

This work suggests various generalizations. First, in \cite{Bertrand:2020nob} we gathered evidence that there 
is a master formulation that universally encodes the conventional 6D, ${\cal N}=(2,2)$ supergravity together with the 
exotic $(3,1)$ and $(4,0)$ theories. Specifically, this master formulation extends  
exceptional field theory, in which fields transform covariantly under $E_{6(6)}$ and depend on $5+27$ coordinates, 
subject to a covariant 
section constraint \cite{Hohm:2013pua,Hohm:2013vpa}. As suggested by the structure of the BPS multiplets, 
this generalization  adds a singlet coordinate and extends  the section constraint whose 
solutions then correspond to  the  $(2,2)$,  $(3,1)$ or $(4,0)$ theory. This master formulation 
is, however, not yet complete for the spin-2 sector, where the analysis of \cite{Bertrand:2020nob} implied 
additional  mass terms for the 5D graviton $h_{\mu\nu}$ that were not visible 
in the $(4,0)$ theory. Remarkably, in terms of the redefined field variables including ${\cal T}_{\mu\nu,\rho\sigma}$ 
precisely such terms emerge in (\ref{L_spin2}), suggesting that this form is much closer to the complete 
master formulation. The main challenge is to find a formulation including ${\cal T}_{\mu\nu,\rho\sigma}$ 
together with a mechanism that renders this field pure gauge for the $(3,1)$ theory which does not feature 
this field.

Most importantly, it remains to actually define the fully interacting version of any of the exotic theories.
We believe that the formulation presented here is particularly promising 
for attempting to write such a non-linear  theory in terms of frame-like field variables. 
The frame formulation of sec.~\ref{subsec:frame} suggests to take the fundamental bosonic fields of the spin-2 
sector to be differential forms with external Lorentz indices, namely: a one-form frame field $E^a=E_{\mu}{}^a dx^{\mu}$ and 
two-forms 
${\cal C}_{(2)}^a=\frac{1}{2}{\cal C}_{\mu\nu}{}^{a} dx^{\mu}\wedge dx^{\nu}$
and ${\cal T}_{(2)}^{ab}=\frac{1}{2}{\cal T}_{\mu\nu}{}^{ab} dx^{\mu}\wedge dx^{\nu}$.  
For instance, the topological terms in the action for ${\cal T}_{(2)}^{ab}$ then take the natural 
Chern-Simons form 
 \be
  {\cal L}_{\rm CS} = -\frac{1}{16} \partial_6{\cal T}_{(2)}{}^{ab}\wedge d{\cal T}_{(2)ab}+\cdots\;, 
 \ee
up to non-linear terms that make the equations of motion compatible with 
$d{\cal T}_{(2)ab}=\partial_6{\cal J}_{(3)ab}$, with the Lorentz-valued 
`3-form current' suggested by the expression (\ref{currentJ}): 
 \be
  {\cal J}_{(3)ab} = \frac{1}{2}\,\varepsilon_{abcde} \left( E^c\wedge {\cal T}_{(2)}^{de} 
  -\tfrac{2}{3} E^c\wedge E^d\wedge E^e\right)+4\,E_{[a}\wedge {\cal C}_{(2)b]}\;. 
 \ee 
Indeed, expanding as $E_{\mu}{}^{a}=\delta_{\mu}{}^{a}+e_{\mu}{}^{a}$ to first order this 
precisely reproduces the current ${\cal J}_{\mu\nu\rho,\alpha\beta}$ used in the main text. 
Similarly, all other terms in the action and symmetry transformations allow for a natural writing 
in terms of differential forms, which in turn suggests at least part of the non-linearities 
that the interacting theory must possess. It remains to define the complete non-linear gauge structure 
and interactions compatible with maximal supersymmetry.

\subsection*{Acknowledgements}

The work of OH is funded   by the European Research Council (ERC) under the European Union's Horizon 2020 research and innovation programme (grant agreement No 771862).

\begin{appendix}

\section*{Appendix}


\section{6D spinor conventions}
\label{app:6D}

\subsection{The gamma matrices}

We are working in 6D space with a mostly positive flat metric $(-,+,+,+,+,+)$. The gamma matrices 
satisfy the following properties
\bea
&&\{\hat{\gamma}_{\hat{\mu}} ,\hat{\gamma}_{\hat{\nu}}\} = 2\eta_{\hat{\mu}\hat{\nu}}\mathbb{I}
\;,\nonumber\\
&&
\hat{\gamma}_{\hat{\mu}}^\dagger = \hat{\gamma}_0\hat{\gamma}_{\hat{\mu}} \hat{\gamma}_0
\;,\nonumber\\
&&
\hat{\gamma}_7^\dagger = \hat{\gamma}_7\;,\quad\mbox{for}\quad
\hat{\gamma}_7 \equiv \gamma_{012346}
\;,\nonumber\\
&&
\hat{\gamma}_7^2 = 1
\;,
\eea
and have been constructed in an explicit basis in (\ref{gamma6}).
These matrices also satisfy a duality relation
\begin{equation}
    \hat{\gamma}^{\hat\mu_1 \cdots \hat\mu_n} = \frac{s_n}{(6-n)!}\, \varepsilon^{\hat\mu_1 \cdots \hat\mu_n \hat\nu_1 \cdots \hat\nu_{6-n}} \hat{\gamma}_{\hat\nu_1 \cdots \hat\nu_{6-n}}\hat{\gamma}_7, \quad \text{where }s_n=
    \begin{cases}
    -1\,:\quad n=0,1,4,5 \\
    +1\,:\quad n=2,3,6
    \end{cases}.
\end{equation}

\subsection{Useful spinor identities}

We are working with chiral spinors and we take the following conventions:
\begin{equation}
    \hat{\gamma}_7{\cal X} = {\cal X}\;,\quad \hat{\gamma}_7\epsilon^A=-\epsilon^A\;,
\end{equation}
for fermionic fields ${\cal X}$ and the supersymmetry parameter $\epsilon^A$, respectively.
This is manifest in the basis (\ref{chiralF}).

The USp(8) indices are raised and lowered using the NW-SE convention i.e. $\epsilon^A = \Omega^{AB}\epsilon_B,\;\epsilon_A = \epsilon^B\Omega_{BA}$ (which entails $\Omega^{AC}\Omega_{CB} = -\delta_B^A$).
Contracting spinors and gamma matrices gives the following identities:
\begin{equation}
    \bar{\epsilon}^A \hat{\gamma}^{(n)}\epsilon^B = t_n \bar{\epsilon}^B \hat{\gamma}^{(n)}\epsilon^A, \quad \text{where }t_n=
    \begin{cases}
    -1\,:\quad n=0,3,4 \\
    +1\,:\quad n=1,2,5,6
    \end{cases},
\end{equation}
where $(n)$ denotes any product of $n$ gamma matrices.
The chirality imposes other conditions:
\begin{align}
    \bar{\epsilon}^A \hat{\gamma}^{(2n+1)} {\cal X} &= 0,\quad \text{(for spinors of opposite chirality)}\;,
    \\
    \bar{\epsilon}^A \hat{\gamma}^{(2n)}\epsilon^B &=0, \quad \text{(for spinors of equal chirality)}
    \;,
    \nonumber
\end{align}
which is proved by inserting a $\gamma_7$ in the contraction. The supersymmetric structure is based on the 
interplay of spinor chirality and (anti)-selfduality of $p$-forms as
\begin{align}
\left(
H_{\hat{\mu}\hat{\nu}\hat{\rho}}{}^{AB}\,
-  \frac{1}{6}\,\varepsilon_{\hat{\mu}\hat{\nu}\hat{\rho}\hat{\sigma}\hat{\tau}\hat{\lambda}} \, H^{\hat{\sigma}\hat{\tau}\hat{\lambda} }{}^{AB}
\right) \hat{\gamma}_{\hat{\mu}\hat{\nu}\hat{\rho}} \epsilon^C =\;& 0
\;.
\end{align}
We also recall the Fierz identity
\begin{equation}
\epsilon_2^A \bar{\epsilon}_1^B = -\frac{1}{4}
\left(\bar{\epsilon}_1^B \hat{\gamma}^{\hat{\mu}} \epsilon_2^A\right)
\hat{\gamma}_{\hat{\mu}} 
+ \frac{1}{48}\left(\bar{\epsilon}_1^B \hat{\gamma}^{\hat{\mu}\hat{\nu}\hat{\rho}} \epsilon_2^A\right) 
\hat{\gamma}_{\hat{\mu}\hat{\nu}\hat{\rho}}
\;.
\end{equation}

\section{USp(8) identities}

Recall, that indices $A, B, \dots$ label the fundamental representation of USp(8).
For an antisymmetric tensor $X^{ABC}=X^{[ABC]}$ we may spell out the traceless projection as
\bea
X^{\llbracket ABC\rrbracket} &=&
X^{ABC} -\frac12\,\Omega^{[AB} X^{C]DE} \Omega_{DE}
\;.
\eea
Similarly, for an antisymmetric and traceless $X^{AB}=X^{\llbracket AB\rrbracket}$ we find
\bea
X^{\llbracket AB} Y^{C\rrbracket} &=&
X^{[AB} Y^{C]} 
+\frac13\,\Omega^{[AB} X^{C]D}Y_D\,.
\eea
For two antisymmetric and traceless tensors $X^{AB}\,,Y^{AB}$, we find
\bea
X^{\llbracket AB} Y^{CD\rrbracket} &=&
X^{[AB} Y^{CD]} + \Omega^{[AB}\,X^{C}{}_E Y^{D]E} -\frac1{12}\,\Omega^{[AB}\Omega^{CD]} X^{EF}Y_{EF}
\;.
\eea
Finally, for antisymmetric and traceless tensors $A_{AB}$, $B_{AB}$, $\phi^{ABCD}$ in the ${\bf 42}$, we have the identity
\bea
A^{CD} B^{E\llbracket A} \phi_{CDE}{}^{B\rrbracket} + B^{CD} A^{E\llbracket A} \phi_{CDE}{}^{B\rrbracket}
&=&
A_{C}{}^{E} B_{ED}\,\phi^{CDAB}
\;.
\eea
This identity is a manifestation of the fact that the tensor product
\bea
({\bf 27} \otimes_{\rm sym} {\bf 27}) \otimes {\bf 42} &=& 
{\bf 1} \oplus {\bf 27} \oplus {\bf 36} \oplus \dots
\;,
\eea
of USp(8) representations only carries a single ${\bf 27}$ representation on the r.h.s.

\end{appendix}


\providecommand{\href}[2]{#2}\begingroup\raggedright\endgroup

\end{document}